\documentstyle[11pt]{article}
\newcommand{\lsim}
 {\ \raise.35ex\hbox{$<$}\kern-0.75em\lower.5ex\hbox{$\sim$}\ }
\newcommand{\gsim}
 {\ \raise.35ex\hbox{$>$}\kern-0.75em\lower.5ex\hbox{$\sim$}\ }
\def\journal #1#2#3#4{#1 {\bf #2} (#4) #3}
\def\PR{Phys.\ Rev.}
\def\PRB{Phys.\ Rev.\ B}
\def\PRL{Phys.\ Rev.\ Lett.}

\def\JPCM{J.\ Phys.\ Cond.\ Mat.}

\def\JPSJ{J.\ Phys.\ Soc.\ Jpn.}

\def\RMP{Rev.\ Mod.\ Phys.}
\def\PTP{Prog.\ Theor.\ Phys.}

\begin{document}
Vers. 19, Jul. 1996
\bigskip\medskip

\centerline
{\Large{\bf
Gutzwiller wave function under magnetic field
}}
\par\bigskip\medskip
\centerline
{\large
Hisatoshi Yokoyama, Shinya Tokizaki
}
\par\bigskip
\centerline
{\it
Department of Physics, Tohoku University, Sendai 980-77, Japan
}
\par\bigskip\bigskip

{\bf Abstract}
\par\noindent
Magnetization process of the Gutzwiller wave function is studied
accurately by a variational Monte Carlo method.
We apply it to the one-dimensional (1D) and 2D Hubbard models (HM),
and to the 1D periodic Anderson model (PAM) without orbital degeneracy.
For the HM, magnetization varies discontinuously to the
full moment, as the magnetic field increases.
For the PAM, the paramagnetic state is unstable against ferromagnetism,
although the energy reduction thereof is small.
\par\bigskip

{\bf keywords}
\par\noindent
magnetization process, Gutzwiller wave function,
Hubbard model, periodic Anderson model
\par\bigskip\bigskip
Author for correspondence
\par\noindent
Name: Hisatoshi Yokoyama
\par\noindent
Postal Address: Department of Physics, Tohoku University,
Aramaki Aoba, Aoba-ku, Sendai, 980-77, Japan
\par\noindent
Fax number: +81-22-217-6447
\par\noindent
E-mail address: yoko@cmpt01.phys.tohoku.ac.jp
\par\bigskip\bigskip
\vfil\eject
\par
Recent experiments under strong magnetic field have renewed
the interests for heavy-electron systems, such as metamagnetic
transitions accompanied by variations of the Fermi surface
and the gap vanishing in the Kondo insulators.
In this paper, with these in mind, we study properties
of the Gutzwiller wave function (GWF)\cite{Gutz} under magnetic field,
in particular magnetization process, applying it to
the Hubbard model (HM) and the periodic Anderson model (PAM).

\par

The GWF with magnetization $m$ is written as
\begin{equation}
|\Psi_\xi(m)\rangle={\cal P}^\ell(\xi)|\Phi_{\rm F}(m)\rangle,
\end{equation}
where $|\Phi_{\rm F}(m)\rangle$ is a spin polarized one-body function
with $m=(N_\uparrow-N_\downarrow)/N_{\rm a}$ ($N_\sigma$ and $N_{\rm a}$
being the numbers of spin $\sigma$ and site, respectively), and
${\cal P}^\ell(\xi)=\prod_i
\left[1-(1-\xi)n^\ell_{i\uparrow}n^\ell_{i\downarrow}\right]$,
$\ell$ being the orbital index, and $\xi$ a variational parameter
which adjusts double occupancy of electrons.
Although the GWF is a primary many-body wave function, most
variational works so far have resorted to the so-called Gutzwiller
approximation (GA)\cite{Gutz,Voll,RU,REH} in estimating variational
expectation values.
Here we use a variational Monte Carlo (VMC) method for exact
evaluation, which tells us how to improve the trial wave function.
\par

First, we discuss the single-band HM,
\begin{equation}
{\cal H}=-t\sum_{<i,j>\sigma}
\left(c_{i\sigma}^{\dagger}c_{j\sigma}+
c_{j\sigma}^{\dagger}c_{i\sigma}\right)
+U\sum_i n_{i\uparrow}n_{i\downarrow}.
\end{equation}
We concentrate on the half-filled band:
$n=(N_\uparrow+N_\downarrow)/N_{\rm a}=1$,
where the magnetic-field effect is the most dominant.
The energy due to the applied field $H$ is given by
a Zeeman term: ${\cal H_{\rm ext}}=-g\mu_{\rm B}H\sum_i S_i^z$.
For this model, $\Phi_{\rm F}(m)$ in Eq.~(1) becomes a simple
Fermi sea, $|\Phi_{\rm F}(m)\rangle=
\prod_{\sigma}\prod_{k\le k_{\rm F}^{\sigma}}
                        c_{k \sigma}^{\dagger} |0\rangle$,
and the effect of magnetic field appears only in magnetization.
To obtain a magnetization curve, first, we minimize the
energy expectation values for zero field, $E_0(m)$, with
respect to the variational parameters ($\xi$ in this case)
for each $m$ by VMC calculations.
Then, we find the relation between $m$ and $H^*$ by determining
$m$ which minimizes $E_{\rm tot}(H^*,m)=E_0(m)-H^*m$ for a given
value of $H^*=g\mu_BH/2$.
\par

Figure 1 shows the magnetization curves in one dimension (1D)
thus obtained for some values of $U/t$ with those of two other
methods.
Under a weak field, these three results agree well.
However, at some critical field $H^*_{\rm c}$, $m$ of
the GWF (and the GA) saturates discontinuously in contrast to the
exact one.
The jump of $m$ becomes large as $U/t$ increases.
The origin of this discontinuity is negative
$\partial^2 E_0(m)/\partial m^2$ for $m\sim 1$, which makes
$E_{\rm tot}$ have double minima for $H^*\sim H^*_{\rm c}$.
A similar feature can be seen for large values of $U/t$
($\gsim 5$) in 2D, as shown in Fig.~2.
For small $U/t$, however, $m$ increases smoothly.
This behavior of $m$ resembles that of the GA\cite{Voll}
except for the existence of the metal-insulator (Brinkman-Rice)
transition at $U=U_{\rm c}$.
If one uses half ellipse as the density of state,
the discontinuity occurs for $U/U_{\rm c}>0.44$.\cite{Voll}
\par

In addition to the 1D exact result,\cite{Takahashi} quantum
Monte Carlo calculations in infinite dimension\cite{SasoHM} concluded
no discontinuity in the magnetization curve.
We consider that this discrepancy is attributed to the onsite nature
of the correlation factor in the GWF; an intersite correlation factor,
spin dependent in this case, will encourage the electron transfer,
which becomes more important as $m$ increases.
\par

Incidentally, for the 1D $t$-$J$ model the GWF does not have a
jump in the magnetization curve for any value of $J/t$ and
well reproduces the exact magnetic properties for the half filling
or the supersymmetric case.\cite{YO1}
This is because the intersite correlation effect is introduced
through the canonical transformation from the HM to the $t$-$J$
model.\cite{YO2}
\par

Having restricted $n$ to the half filling, we have confirmed that
the tendency mentioned here holds also for $n<1$.
\par

Next, we consider the 1D periodic Anderson model without f-level
degeneracy,
\begin{eqnarray}
{\cal H}&=&\sum_{k\sigma}\varepsilon_k c^\dagger_{k\sigma}c_{k\sigma}
          -\sum_{k\sigma}V_k\left(
c^\dagger_{k\sigma}f_{k\sigma}+f^\dagger_{k\sigma}c_{k\sigma}\right)
        \nonumber\\
      &&+\sum_{k\sigma}\varepsilon_{\rm f}f^\dagger_{k\sigma}f_{k\sigma}
        +U\sum_j n^{\rm f}_{j\uparrow} n^{\rm f}_{j\downarrow}.
\end{eqnarray}
For simplicity, we put $\varepsilon_k=-2t\cos k$,
$\varepsilon_{\rm f}$ and $V_k(\equiv V)$ constant, and fix two
parameters at typical and convenient values: $V/t=0.5$ and $U/t=\infty$.
Furthermore, as for the Zeeman term we assume $g_{\rm c}=g_{\rm f}$.
\par

For this Hamiltonian, we consider the Gutzwiller-projected hybridized
band state: Eq.~(1) with ${\cal P}^{\rm f}(0)$ and
\begin{equation}
|\Phi_{\rm F}(m)\rangle=\prod_{\sigma}\prod_{k\le k_{\rm F}^{\sigma}}
    \left[\cos\phi_k c_{k \sigma}^{\dagger}
    +\sin\phi_k f_{k \sigma}^{\dagger}\right]|0\rangle,
\end{equation}
where we choose $\phi_k$ as a spin-independent noninteracting form:
\begin{equation}
\tan\phi_k=\frac{2\tilde V}{\tilde\varepsilon_{\rm f}-\varepsilon_k
\pm\sqrt{\left(\tilde\varepsilon_{\rm f}-\varepsilon_k\right)^2
+4\tilde V^2}}
\end{equation}
for the upper ($-$) and lower ($+$) bands, respectively.
$\tilde\varepsilon_{\rm f}$ and $\tilde V$ are variational
parameters, which adjust the band form.
This wave function is an extention of Yosida's singlet
state for the Kondo problem,\cite{Yosida} and has been
extensively studied for the cases without magnetic field.\cite{Shiba}
\par

In Fig.~3 the energy expectation values without field are
depicted.
We have calculated for metallic ($n=1.6$) as well as insulating
($n=2.0$) cases.
In each case, $E_0(m)/t$ varies slightly
for small $m$ and then increases abruptly with increasing $m$.
This aspect is similar to the exact-diagonalization results.\cite{Ueda1}
Meanwhile, the optimized state has finite spin polarization,
as well as $\partial E_0(m)/\partial m|_{m=0}>0$ and
$\partial^2 E_0(m)/\partial m^2|_{m=0}<0$.
Also for a finite $U$ ($U,V,E_{\rm f}=1.0t,0.5t,-1.0t$),
we have confirmed that the above derivatives have the same
signs with those for infinite $U$; namely the GWF is unstable
against a magnetic order.
These aspects agree with the Gutzwiller approximations,\cite{RU,REH}
but does not with an exact analysis for the symmetric case,\cite{Ueda2}
and a study in infinite dimension,\cite{SasoPAM} where the ground state
is paramagnetic.
>From these facts we observe that the ground state of the 1D
nondegenerate PAM, either paramagnetic or partially ferromagnetic,
is rather fragile;
actually the energy differences in Fig.~3 are the order of $0.01t$.
Stable paramagnetism may need f-level degeneracy.\cite{RU}
\par

Magnetization curves obtained from Fig.~3 are shown in Fig.~4.
There is no sign of metamagnetism.
It also remains a future problem how the gap for the insulating case
changes as a function of various parameters.
\par

In conclusion, although the VMC results support the GA qualitatively,
improvements of the trial function are needed for advanced
discussions.
To this end, introduction of a spin-dependent one-body part
as well as of an intersite (RKKY-like) or off-diagonal correlation
factors is important.
\par

\par
\bigskip\bigskip
{\bf Figure captions}
\par\bigskip\noindent
{\bf Fig.~1}\
Magnetization as a function of magnetic field for the 1D Hubbard model.
The results of the GWF are shown with those of the GA and the Bethe
Ansatz (Exact)\cite{Takahashi} for $U/t=3$.
The system used for the GWF has 90 sites with closed-shell
condition.
For the GA the 1D cosine band is assumed.

\par\bigskip\noindent
{\bf Fig.~2}\
Magnetization as a function of magnetic field of the Hubbard model
for the 2D square lattice by the GWF.
We use systems of $L\times L$ ($L=10,12,14$) sites with
the antiperiodic-antiperiodic boundary condition and closed shell
condition.

\par\bigskip\noindent
{\bf Fig.~3}\
Energy expectation values without magnetic field as a function of
magnetization for 1D periodic Anderson model.
Minimal values for $E_{\rm f}/t=-1.0$ are plotted by dotted
lineas as a guide.
The system of $N_{\rm a}=50$ is used with closed shell condition.

\par\bigskip\noindent
{\bf Fig.~4}\
Magnetization as a function of magnetic field of the 1D periodic
Anderson model for some parameter values.
For comparison, the data for $U/t=0, V/t=0.5$ and $E_{\rm f}/t=-1.0$
are plotted by dash-dotted lines.

\end{document}